# Generating controllable Laguerre-Gaussian laser modes through intracavity spin-orbital angular momentum conversion of light

Dunzhao Wei, [1] Yue Cheng, [1] Rui Ni, [1] Yong Zhang,[1,*] Xiaopeng Hu,[1,†] Shining Zhu, [1] and Min Xiao[1,2,‡]

[1]National Laboratory of Solid State Microstructures, College of Engineering and Applied Sciences, and School of Physics, Nanjing University, Nanjing 210093, China
[2]Department of Physics, University of Arkansas, Fayetteville, Arkansas 72701, USA

**Abstract**

The rapid developments in orbital-angular-momentum-carrying Laguerre-Gaussian ($LG_l^0$) modes in recent years have facilitated progresses in optical communication, micromanipulation and quantum information. However, it is still challenging to efficiently generate bright, pure and selectable $LG_l^0$ laser modes in compact devices. Here, we demonstrate a low-threshold solid-state laser that can directly output selected high-purity $LG_l^0$ modes with high efficiency and controllability. Spin-orbital angular momentum conversion of light is used to reversibly convert the transverse modes inside cavity and determine the output mode index. The generated $LG_1^0$ and $LG_2^0$ laser modes have purities of ~97% and ~93% and slope efficiencies of ~11% and ~5.1%, respectively. Moreover, our cavity design can be easily extended to produce higher-order Laguerre-Gaussian modes and cylindrical vector beams. Such compact laser configuration features flexible control, low threshold, and robustness, making it a practical tool for applications in super-resolution imaging, high-precision interferometer and quantum correlations.

*zhangyong@nju.edu.cn
†xphu@nju.edu.cn
‡mxiao@uark.edu

**Introduction**

Laguerre-Gaussian ($LG_l^p$) modes, characterized by the azimuthal index *l* (any integers) and the radial index *p* (zero or positive integers), are the eigen solutions of the paraxial wave equation in cylindrical coordinates[1]. These $LG_l^p$ modes form a complete and orthonormal set, so that an arbitrary spatial light field can be expressed as the superposition of $LG_l^p$ modes[2]. The $LG_l^p$ modes with $p = 0$, as special optical vortexes, have been a hot research topic since the orbital angular momentum (OAM) properties of light beam have been identified in the azimuthal phase term $e^{il\varphi}$ of the $LG_l^0$ modes[3]. The $LG_l^0$ mode carries an OAM of $l\hbar$ per photon, where $\hbar$ is the reduced Planck constant and $l$ is also called topological charge (TC). Featuring donut-shaped intensity distributions and carrying OAM, the $LG_l^0$ modes have been widely applied in optical tweezers, nonlinear and quantum optics, optical communication, super-resolution imaging, material processing, rotation Doppler effect, and so on[4-15]. The evolving practical applications inevitably require high-quality $LG_l^p$ laser modes. For example, a pure $LG_l^0$ mode has a more uniform donut-intensity distribution along the propagation direction, which can increase the depth resolution in super-resolution imaging[16]. In the field of precision measurement, it has been shown that higher-order $LG_l^p$ modes could reduce thermal noise in the LIGO system for gravitational-wave detection because of their more homogeneous power distributions[17,18]. The high-precision optical interferometry in the LIGO system requires a high-purity $LG_l^p$ mode. In quantum information science, high-purity $LG_l^p$ modes could increase hybrid azimuthal-radial quantum correlations[19-21]. However, it is still a challenge to efficiently generate high-purity $LG_l^p$ modes in a simple optical setting. In this Letter, we experimentally demonstrate an intracavity spin-orbital angular momentum conversion method to produce selected high-purity $LG_l^0$ laser modes, which can be feasibly extended to generate a general $LG_l^p$ laser mode.

The common way to generate a $LG_l^0$ beam is to add the azimuthal phase term of $e^{il\varphi}$ to a Gaussian mode using a fork-grating, a q-plate, a spiral phase plate, and so on[22]. However, directly adding a spiral phase to the Gaussian mode will generate a hypergeometric-Gaussian (HyGG) mode, which has a definite $l$ index but an expansion of the $p$ index[23,24], which decreases the mode purity and conversion efficiency of the desired $LG_l^0$ beam. Furthermore, the power weightings of these undesired $LG_l^p$ (with $p > 0$) modes have positive correlations to the modulus of the $l$ index. These undesired radial modes hidden in the $LG_l^0$ beam will decrease the imaging resolution in the super-resolution microscope, reduce the accuracy of the high precision interferometer[25], contaminate the quantum information processing[19,20], and induce uncertainties in the nonlinear interactions[26].

Researchers have turned to intracavity modulation for generating high-purity $LG_l^0$ modes, since an optical cavity can perform mode selection and directly output laser modes[27,28]. Unfortunately, because $LG_l^0$ modes with opposite handedness (for example, $l = 1$ and $l = -1$ modes) have the same loss and are degenerate in an active optical cavity, their coherent superpositions usually form petal-intensity patterns[29,30]. The superposition of two petal-intensity patterns may form a donut-intensity profile but not a pure $LG_l^0$ mode[31]. Special components, like nanoscale stripes and oblique etalons, are proposed to discriminate the degenerate $LG_l^0$ modes inside active cavities. However, the additional components with circular asymmetries spoil the mode purities of the generated $LG_l^0$ modes, which has impeded their further development[32,33]. Recently, q-plates, a kind of planar phase-modulation devices, is utilized inside the laser cavity for generating high-purity $LG_l^0$ modes[34]. However, the extremely high pump power needed for operation severely limits its practical applications[34]. To generate useful $LG_l^0$ laser modes with high mode purities, low lasing thresholds and controllable mode indices, the laser cavity needs to be carefully designed with proper control and selection of intracavity components.

In our experiment, we put a vortex half-wave plate (VWP, Thorlabs, Inc.), a Faraday rotator (FR) and a quarter-wave plate (QWP) inside a Nd: YVO$_4$ laser cavity, which can achieve cavity mode reversibility through intracavity spin-orbital angular momentum conversion. The process is self-producing after each round trip. The QWP generates a circularly-polarized state carrying a spin angular momentum (SAM) of $\pm\hbar$ per photon depending on its handedness[35]. The interaction between a photon and an optically anisotropic medium will change the value of the photon's SAM and induce a geometric phase shift[36]. The VWP, featuring the artificial helical anisotropic parameter space, can add a spiral geometric phase to an incident mode by flipping the handedness of its circular polarization, *i.e.*, spin-orbital angular momentum conversion. Therefore, the VWP couples the SAM or polarization of the intracavity field into the OAM, and flexibly controls the TC[37]. The modulus and sign of the TC can be changed by using different VWPs and rotating the QWP, respectively. By adding a pinhole to suppress the undesired radial modes, the laser outputs a selected high-purity $LG_l^0$ laser mode for example, a $LG_1^0 (LG_2^0)$ mode with a mode purity of ~97% (~93%), at the wavelength of 1064 nm in our experiment. The specially-designed compact cavity requires only a few intracavity components, which benefits the system stability, loss control and practical operation. The lasing thresholds for lower order $LG_l^0$ modes are comparable to that of the Gaussian mode and the slope efficiencies of the $LG_1^0$ and $LG_2^0$ lasing modes are ~11% and ~5.1%, respectively. The beam quality factor and mode stability at high pump power also show the excellent laser performance. The flexibility of the cavity design is also reflected in its ability to generate laser outputs of vector beams with cylindrical symmetry in polarization by slightly modifying the intracavity elements[38].

## Materials and Methods

**Principle of intracavity spin-orbital angular momentum conversion.**

In a stable laser cavity, a light beam should reproduce itself after each round trip when the cavity resonates. Since each point on the VWP is a half-wave retarder, a linearly-polarized light beam, propagating forward and then reflecting backward through the same VWP, is equivalent to experience a full-wave plate, in which the initial polarization and wavefront are recovered (Fig. 1a). To couple the spin and orbital angular momenta, a QWP is needed to turn the linearly-polarized state to the circularly-polarized state (see Supplementary Note 1 for the working principle of the VWP). The combination of the FR and QWP makes the light beam reversible when its initial polarization direction is parallel to the optical axis of the QWP (Fig. 1b). Therefore, combining the FR, QWP and VWP, we achieve a complete reversible cycle inside a laser cavity for the output of a selected OAM-carrying mode (Fig. 1c). Here, the FR enables mode reversible propagation. The system is greatly simplified in comparison to the previous intracavity geometric-phase-control configuration[34]. Figure 1d shows a one-round-trip mode transformation for laser output of a $|l_0+m, R\rangle$ mode in the configuration shown in Fig. 1c. The configuration performs the transformation $e^{il_0\varphi}|H\rangle \to e^{i(l_0+m)\varphi}|R\rangle$ in the forward propagation and $e^{-i(l_0+m)\varphi}|L\rangle \to e^{-il_0\varphi}|H\rangle$ in the backward propagation, where $H$, $R$, and $L$ refer to horizontally-polarized, left-circularly-polarized (LCP), and right-circularly-polarized (RCP) states, respectively, $l_0$ refers to the TC of the input mode, and the positive integer $m$ is the additional TC determined by the VWP. The intracavity spin-orbital angular momentum conversion happens when light passing through the VWP, i.e. $e^{il_0\varphi}|L\rangle \to e^{i(l_0+m)\varphi}|R\rangle$ for forward conversion and $e^{-i(l_0+m)\varphi}|L\rangle \to e^{-il_0\varphi}|R\rangle$ for backward conversion, considering that the reflection at output coupler induces the inversions of handedness for both SAM and OAM. The additional TC of $m$ is cancelled in a round trip to satisfy the reversible transformation in the cavity. In addition, orienting the fast axis of the QWP vertically, the forward conversion becomes $e^{il_0\varphi}|H\rangle \to e^{i(l_0-m)\varphi}|L\rangle$, which will output the $|l_0-m, L\rangle$ mode. In our experiment, the initial mode carries a TC of $l_0 = 0$, so the TC of an output mode is directly controlled by the QWP and the VWP (see Supplementary Note 2 for the self-reproducing process).

**Laser Cavity design.**

Figure 2a shows the experimental setup. An end-pumped solid-state laser, working in continuous-wave mode at 1064 nm wavelength, is comprised of a Nd: YVO$_4$ crystal as the gain medium, a film coated on the front face of the crystal as an input coupler, a lens with a focal length $f$, and a partially transmitted plane mirror as an output coupler. The FR, QWP and VWP are inserted into this laser cavity to achieve reversible mode conversion during spin-orbital coupling. The lens divides the cavity into two parts (Part 1 and Part 2 in Fig. 2a) with lengths labelled as $L_A$ and $L_B$, satisfying the cavity stability condition[39]

$$0 \leq \left(1-\frac{L_A}{f}\right)\left(1-\frac{L_B}{f}\right) \leq 1. \qquad (1)$$

We choose to work under the condition of $L_A < f$ and $L_B < f$. Considering the refractive indices of the gain medium, FR and QWP in the first part of the cavity, the actual $L_A$ can be a bit larger than $f$ without breaking the cavity stability condition. To optimize mode matching, $L_A$ is set to approach $f$ so that the cavity mode has a diameter large enough at the position of the lens and a diameter comparable to that of the pump beam at the position of the gain medium. The pump beam is focused on the crystal with a 200 um-in-diameter Gaussian spot. The VWP is placed in Part 2 of the cavity, adjacent to the lens to take advantage of the large mode size, because the large mode size can effectively overcome the imperfect singularity of the VWP. A polarization beam splitter (PBS) is inserted between the gain medium and the FR to eliminate the unwanted vertically-polarized light caused by the imperfect polarization modulation from the FR, QWP and VWP. A 1 mm-in-diameter pinhole is placed at the front of the output coupler to filter out the higher-order transverse modes so that the cavity can output the desired $LG_l^0$ beam. Therefore, in an ideal case with a highly-efficient intracavity mode conversion, a $LG_0^0$ mode is expected to oscillate in Part 1 of the cavity, while a $LG_l^0$ mode oscillates in the regime of Part 2. Under such experimental configuration, higher-$p$-order LG modes are greatly suppressed and the laser will output a selected high-purity circularly-polarized $LG_l^0$ mode. Figure 2b shows an output $LG_1^0$ laser mode generated by the VWP of $m = 1$. The $LG_{-1}^0$ mode can be generated (Fig. 2c) by rotating the QWP with an angle of 90°. Replacing the VWP with $m = 2$, we have generated the $LG_2^0$ (or $LG_{-2}^0$) mode by orienting the fast axis of the QWP horizontally (or vertically) as shown in Fig. 2d (or Fig. 2e). All the intensity profiles present high-quality donut shape. Simulations based on Fox-Li method have been carried out to find the steady cavity mode and the corresponding output mode for each cavity configuration[40], which validate our experimental setup (see Supplementary Note 6 for the numerical simulation). Since the experimental setup is similar for generating RCP $LG_{|l|}^0$ and LCP $LG_{-|l|}^0$ modes, we will only present the laser properties of RCP $LG_1^0$ and $LG_2^0$ laser modes in the following.

**Detailed parameters of the experimental setup.**

As shown in Fig. 2a, an a-cut 0.5 at% Nd-doped YVO$_4$ crystal with its dimensions of 3 mm × 3 mm × 8 mm serves as the gain medium. One of its ends, as the input coupler, is coated with the reflectivity >99.9% at 1064 nm and transmittance >97.7% at 808 nm; the other end has a high transmittance of >99.8% at 1064 nm. The crystal is mounted on a water-cooled copper holder maintained at the room temperature and is oriented to generate horizontally-polarized light. The output coupler is a plane mirror with a transmittance of 10% at 1064 nm. The lens with a focal length $f$ is inserted into the cavity to keep the cavity satisfying the stability condition. The pump source is a fiber-coupled laser diode outputting a continuous-wave at 808 nm wavelength, which is coupled out by a telescope and forms a 200 μm-in-diameter focusing spot in the crystal. The FR, QWP and VWP are inserted into the laser cavity to achieve reversible mode conversion inside the cavity. The PBS next to the crystal eliminates unwanted vertically-polarized light and the pinhole with a diameter of 1 mm suppresses the higher-$p$-order modes. To generate $LG_1^0/LG_{-1}^0$ (or $LG_2^0/LG_{-2}^0$) laser modes, we use the VWP of $m = 1$ (or $m = 2$). The cavity parameters are set to be $f = 200$ mm, $L_A = 205$ mm and $L_B = 45$ mm for $LG_1^0/LG_{-1}^0$ modes and $f = 400$ mm, $L_A = 380$ mm and $L_B = 45$ mm for $LG_2^0/LG_{-2}^0$ modes. The output laser beams are recorded by a laser beam profiler (LBP, Newport Corporation).

## Results and Discussion

**Lasing threshold and slope efficiency.**

Figure 3 shows the output power dependences on the pump power for the generations of $LG_0^0$, $LG_1^0$ and $LG_2^0$ laser modes. When the VWP is absent, the laser outputs a LCP $LG_0^0$ mode (i.e., Gaussian mode) with a lasing threshold of ~0.5 W and a slope efficiency of 18.2%. Despite the extra loss induced by inserting the VWP, the threshold of the lasing $LG_1^0$ mode is only ~0.7 W with a corresponding slope efficiency of 11.06%. When the pump power increases to 1.7 W, the output power of the $LG_1^0$ mode reaches 120 mW. The lasing threshold is increased to 1.8 W for the $LG_2^0$ mode generation with a slope efficiency of 5.11%, because the loss due to the undesired modes generated by the VWP increases with the $l$ value. Under the present experimental setup, the output power is limited by the damage thresholds of the components in the cavity. The intensity profiles are preserved well under different pump powers (see Supplementary Fig. 1). For generating high-power and/or higher-order LG modes, it is crucial to further improve the performances of the intracavity components, especially the conversion efficiency of the VWP.

**Mode purities.**

Next, we present the transverse-mode-selection ability of the laser cavity by measuring the mode purities of the output $LG_l^0$ modes. We use modal decomposition method to measure the purities of the output $LG_l^0$ laser modes under the output power of ~40 mW (see Supplementary Note 4 for the modal decomposition process) [24,41]. A group of pure $LG_l^p$ modes with $p = 0$ to 5 and $l = 0, 1, 2$ are chosen to decompose the output $LG_1^0$ laser mode to determine its mode purity. Figure 4a shows the decomposing results in a histogram. The $x$ axis and $y$ axis represent the $p$ index and $l$ index, respectively, while the $z$ axis shows the power weightings of different $LG_l^p$ modes. The measured purity of the output $LG_1^0$ mode is 96.8 %. Other undesired modes have been suppressed to near zero. To measure the output $LG_2^0$ laser mode, we choose a group of pure $LG_l^p$ modes with $p = 0$ to 5 and $l = 1, 2, 3$. The result shows the main power weighting of $LG_2^0$ mode to be 92.7% (Fig. 4b). The active cavity plays an important role in the transverse-mode selection, which increases the mode purities appreciably, as compared to the achieved mode purities through extracavity modulation using VWPs (typically ~80% for $LG_1^0$ mode and ~60% for $LG_2^0$ mode)[24].

**Beam quality factors.**

We further measure the beam quality factor to characterize the output $LG_l^0$ laser mode. The output power is set to be ~40 mW. We record output intensity patterns of the $LG_1^0$ and $LG_2^0$ modes (Figs. 5a and 5b) at different propagation distances separately. The corresponding distances are 20 cm, 30 cm, 40 cm, 50 cm and 60 cm from left to right in Fig. 5, respectively. The intensity patterns preserve the well-defined donut shape with negligible higher-order radial modes during the propagation. The diameters of the intensity patterns increase due to the divergence of laser beams. The measured beam quality factors are 2.19 for the $LG_1^0$ mode and 3.74 for the $LG_2^0$ mode. For an ideal $LG_l^p$ mode, the beam quality factor should be $2p+|l|+1$, so the corresponding quality factors are 2 and 3 for ideal $LG_1^0$ and $LG_2^0$ modes, respectively. Our experimental results show the good beam quality of the generated $LG_l^0$ modes.

**Laser output of cylindrical vector beams.**

Laser beams with different axial symmetries in polarization, the so-called cylindrical vector beams, can also be generated by different combinations of VWPs and linearly-polarized directions of incident beams[38,42]. As shown in Fig. 1(a), a single VWP can achieve a reversible propagation cycle inside a laser cavity, which can be used to directly produce laser outputs of cylindrical vector beams. The new setup satisfies the self-reproducing and stability conditions for a cavity (see Supplementary Fig. 4). The light beam incident on the VWP is fixed into a horizontal polarization in our experiment, so the output polarization distribution can be directly controlled by rotating the VWP (See Supplementary Equation (3)). A linear polarizer is used to analyze the polarization distribution of the output beam. Figures 6a and 6b show the output cylindrical vector beams with a well-defined donut-intensity distribution by orienting the VWP of $m = 1$ with different initial angles of $\varphi_0 = 45°$ and $\varphi_0 = 0°$, respectively. By orienting the linear polarizer at the angles of 0, 45°, 90°, and 135°, the measured intensity distributions are shown in Figs. 6d and 6e, which indicate that the output beams shown in Figs. 6a and 6b are azimuthal and radial vector beams, respectively[43]. Using the VWP of $m = 2$, we have generated a higher-order cylindrical vector beam with negligible radial modes (Fig. 6c). The measured intensity distributions after the linear polarizer, as shown in Fig. 6f, reveal the cylindrical polarization distribution of Fig. 6c. The cylindrical vector beams with a well-define donut-intensity profile and controllable polarization distribution can be applied in optical communication, material processing, superfocusing, etc[43-46].

**Conclusions**

In summary, we have demonstrated a Nd: YVO$_4$ laser generating high-purity $LG_l^0$ laser modes. The cavity design satisfies the light reversibility and stability conditions for a cavity. The degeneracy of opposite-handedness LG modes is lifted through coupling SAM into OAM for the field mode in the laser cavity. The output power dependences on the pump power show low lasing thresholds (0.7 W for $LG_1^0$ and 1.8 W for $LG_2^0$) and high slope efficiencies (11.06% for $LG_1^0$ and 5.11% for $LG_2^0$) for the generated $LG_l^0$ modes. The mode purities are up to ~97% and ~93% for the output $LG_1^0$ and $LG_2^0$ modes, respectively. Changing the fast-axis orientation of the QWP, the generated LCP $LG_{-1}^0$ and $LG_{-2}^0$ modes have similar high-quality features as those of the $LG_1^0$ and $LG_2^0$ modes. The intensity profiles remain stable at different pump powers showing the mode stability of the cavity. Those features make the $LG_l^0$ laser beam practical and preferred choice for applications in optical tweezers, optical metrology, super-resolution imaging, material processing, and so on. By further simplifying the cavity, different high-quality cylindrical vector beams have also been produced, which can be used in superfocusing and particle trapping. Less complexity in cavity construction and adjustment will allow our scheme to be easily transplanted into other laser and nonlinear systems, such as Ti: Sapphire lasers, He-Ne lasers, CO$_2$ lasers, and even optical parametric oscillators. Furthermore, due to the flexible controllability of the azimuthal index and the low lasing threshold, it is possible for using such setup to generate high-quality $LG_l^p$ modes by replacing the VWP with $LG_l^p$ modal q-plates or J-plates[47,48], which can be used to further increase the precision of gravitational-wave detection, extend high-dimensional entanglement and optical communication bandwidth, and design more sophisticated trapping configurations[19,20,25,49,50].

**Acknowledgements**

This work was supported by National Key R&D Program of China (2016YFA0302500, 2017YFA0303700); National Natural Science Foundation of China (NSFC) (91636106, 11621091).

**Conflict of Interest:**

The authors declare no competing financial interests.

**Author contributions**

D.W., Y.C., and R.N. performed the experiments and simulations under the guidance of Y.Z., X.H. S.Z., and M.X. D.W., Y.Z., and M.X. wrote the manuscript with contributions from all co-authors.

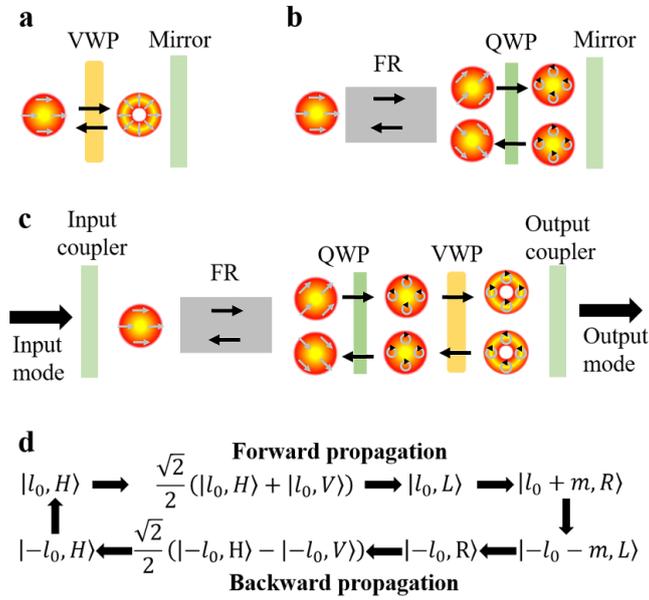

**Figure 1. Reversible light transmission in a laser cavity.** (a) A reversible optical setup including a VWP and a mirror. (b) A reversible optical setup including a FR, a QWP, and a mirror. (c) A reversible optical cavity combining the FR, QWP and VWP for OAM mode generation. (d) The complete reversible cycle of mode transformation in (c).

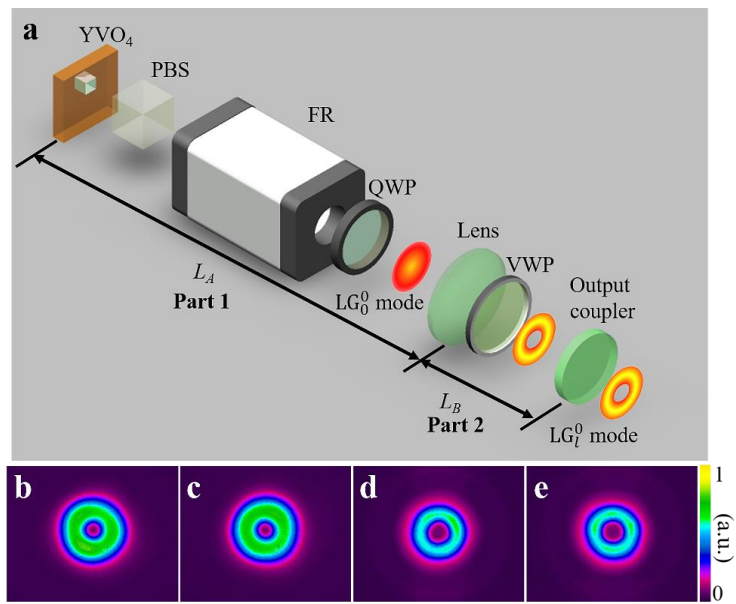

**Figure 2. Experimental setup and the output intensity patterns.** (a) The experimental setup for generating $LG_l^0$ laser modes. An end-pumped solid-state laser comprises of a Nd: YVO$_4$ crystal, a lens and an output coupler working at the wavelength of 1064 nm. The crystal is pumped by an 808 nm-wavelength fiber-coupled diode laser. Following the crystal, a polarization beam splitter (PBS), a Faraday Rotator (FR), a quarter-wave plate (QWP) and a vortex half-wave plate (VWP) are placed in the cavity successively for polarization state control and spin-orbital angular momentum coupling. All of them have antireflection coatings and are oriented precisely, so that the whole cavity satisfies the reversible propagation condition. (b) and (c) are the output $LG_1^0$ and $LG_{-1}^0$ modes using the VWP of $m = 1$, respectively. (d) and (e) are the output $LG_2^0$ and $LG_{-2}^0$ modes using the VWP of $m = 2$, respectively.

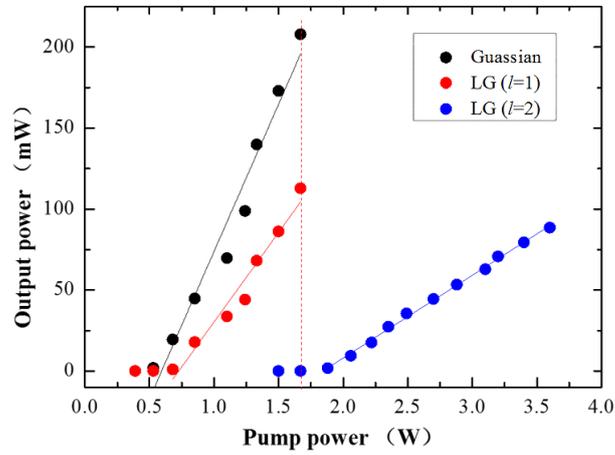

**Figure 3. Output powers versus the pump power.** The experimentally measured output powers of the $LG_0^0$, $LG_1^0$ and $LG_2^0$ laser modes are shown in black, red and blue dots, respectively. The corresponding fitting results are shown in color lines.

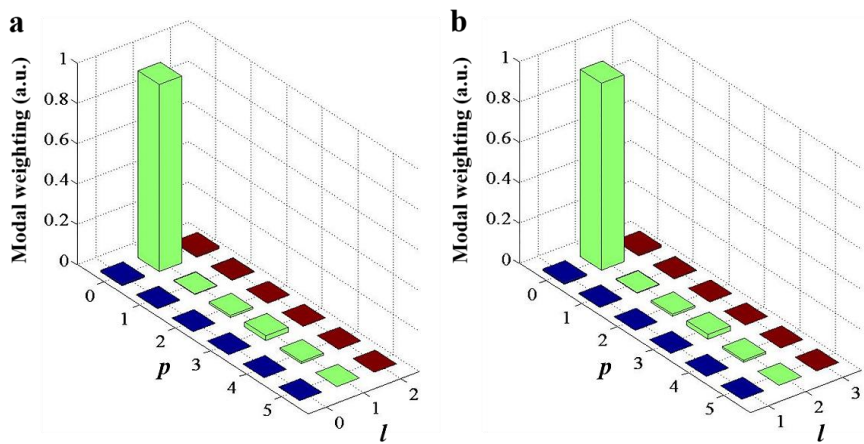

**Figure 4. Modal decomposition results**. The modal decomposition results of the output $LG_1^0$ mode (a) and $LG_2^0$ mode (b), respectively.

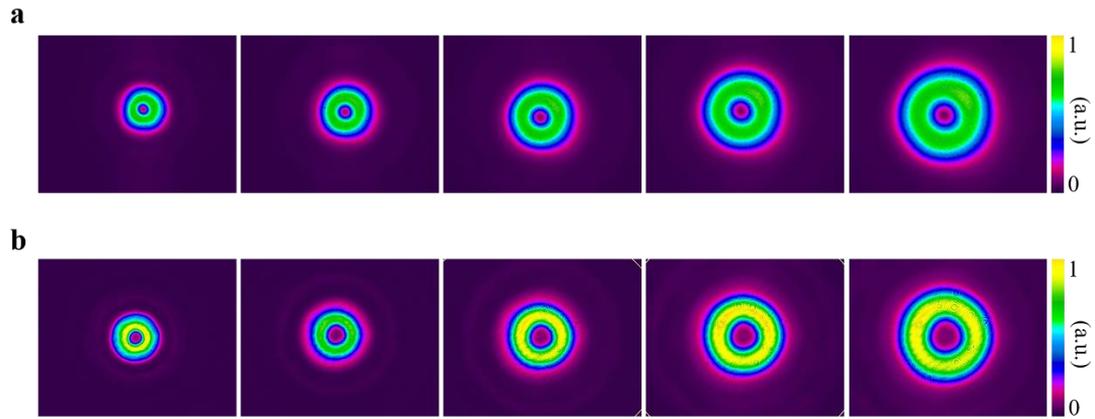

**Figure 5. Intensity patterns of the propagating $LG_l^0$ laser beams.** Intensity patterns of the output $LG_1^0$ mode (a) and $LG_2^0$ mode (b) at propagation distances of 20 cm, 30 cm, 40 cm, 50 cm, and 60 cm from left to right, respectively.

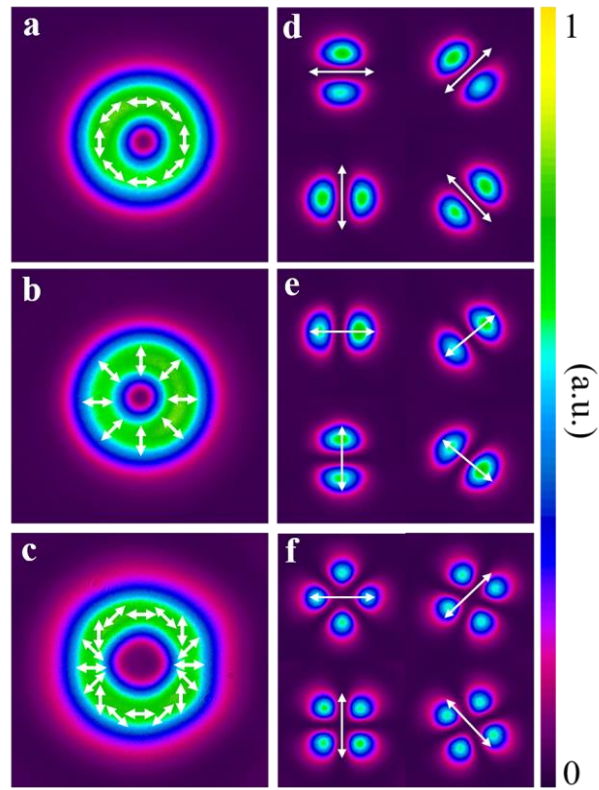

**Figure 6. Intensity patterns of cylindrical vector laser beams.** Intensity patterns of the azimuthal (a) and radial (b) vector beams generated by the VWP of $m$ = 1. (c) A higher-order cylindrical vector beam generated by the VWP of $m$ = 2. (d) to (f) are the measured intensity distributions after passing

through a linear polarizer oriented in different angles of 0, 45°, 90° and 135°, which correspond to (a) to (c) respectively. Arrows are used to indicate the polarization distributions in (a) to (c) and the orientations of the polarizer in (d) to (f).

**References**

1   Kogelnik, H. & Li, T. Laser Beams and Resonators. *Appl. Opt.* **5**, 1550-1567 (1966).
2   Forbes, A., Dudley, A. & McLaren, M. Creation and detection of optical modes with spatial light modulators. *Adv. Opt. Photonics* **8**, 200-227 (2016).
3   Allen, L., Beijersbergen, M. W., Spreeuw, R. J. & Woerdman, J. P. Orbital angular momentum of light and the transformation of Laguerre-Gaussian laser modes. *Phys. Rev. A* **45**, 8185-8189 (1992).
4   Simpson, N. B., Dholakia, K., Allen, L. & Padgett, M. J. Mechanical equivalence of spin and orbital angular momentum of light: an optical spanner. *Opt. Lett.* **22**, 52-54 (1997).
5   Mair, A., Vaziri, A., Weihs, G. & Zeilinger, A. Entanglement of the orbital angular momentum states of photons. *Nature* **412**, 313-316 (2001).
6   Wei, D. *et al.* Directly generating orbital angular momentum in second-harmonic waves with a spirally poled nonlinear photonic crystal. *Appl. Phys. Lett.* **110**, 261104 (2017).
7   Fang, X. *et al.* Coupled orbital angular momentum conversions in a quasi-periodically poled LiTaO3 crystal. *Opt. Lett.* **41**, 1169-1172 (2016).
8   Nicolas, A. *et al.* A quantum memory for orbital angular momentum photonic qubits. *Nature Photon.* **8**, 234-238 (2014).
9   Bozinovic, N. *et al.* Terabit-scale orbital angular momentum mode division multiplexing in fibers. *Science* **340**, 1545-1548 (2013).
10  Wang, J. *et al.* Terabit free-space data transmission employing orbital angular momentum multiplexing. *Nature Photon.* **6**, 488-496 (2012).
11  Hell, S. W. & Wichmann, J. Breaking the diffraction resolution limit by stimulated emission: stimulated-emission-depletion fluorescence microscopy. *Opt. Lett.* **19**, 780-782 (1994).
12  Alexandrescu, A., Cojoc, D. & Di Fabrizio, E. Mechanism of angular momentum exchange between molecules and Laguerre-Gaussian beams. *Phys. Rev. Lett.* **96**, 243001 (2006).
13  Hamazaki, J. *et al.* Optical-vortex laser ablation. *Opt. Express* **18**, 2144-2151 (2010).
14  Arlt, J. *et al.* Moving interference patterns created using the angular Doppler-effect. *Opt. Express* **10**, 844-852 (2002).
15  Ni, J. *et al.* Three-dimensional chiral microstructures fabricated by structured optical vortices in isotropic material. *Light Sci. Appl.* **6**, e17011 (2017).
16  Yu, W. *et al.* Super-resolution deep imaging with hollow Bessel beam STED microscopy. *Laser Photonics Rev.* **10**, 147-152 (2016).
17  Mours, B., Tournefier, E. & Vinet, J.-Y. Thermal noise reduction in interferometric gravitational wave antennas: using high order TEM modes. *Class. Quant. Grav.* **23**, 5777-5784 (2006).
18  Chelkowski, S., Hild, S. & Freise, A. Prospects of higher-order Laguerre-Gauss modes in future

gravitational wave detectors. *Phys. Rev. D* **79**, 122002 (2009).

19  Salakhutdinov, V. D., Eliel, E. R. & Loffler, W. Full-field quantum correlations of spatially entangled photons. *Phys. Rev. Lett.* **108**, 173604 (2012).

20  Karimi, E. *et al.* Exploring the quantum nature of the radial degree of freedom of a photon via Hong-Ou-Mandel interference. *Phys. Rev. A* **89**, 013829 (2014).

21  Krenn, M., Malik, M., Erhard, M. & Zeilinger, A. Orbital angular momentum of photons and the entanglement of Laguerre-Gaussian modes. *Phil. Trans. R. Soc. A* **375**, 1-21 (2017).

22  Yao, A. M. & Padgett, M. J. Orbital angular momentum: origins, behavior and applications. *Adv. Opt. Photonics* **3**, 161-204 (2011).

23  Karimi, E., Zito, G., Piccirillo, B., Marrucci, L. & Santamato, E. Hypergeometric-gaussian modes. *Opt. Lett.* **32**, 3053-3055 (2007).

24  Sephton, B., Dudley, A. & Forbes, A. Revealing the radial modes in vortex beams. *Appl. Opt.* **55**, 7830-7835 (2016).

25  Granata, M., Buy, C., Ward, R. & Barsuglia, M. Higher-order Laguerre-Gauss mode generation and interferometry for gravitational wave detectors. *Phys. Rev. Lett.* **105**, 231102 (2010).

26  Lanning, R. N. *et al.* Gaussian-beam-propagation theory for nonlinear optics involving an analytical treatment of orbital-angular-momentum transfer. *Phys. Rev. A* **96**, 013830 (2017).

27  Senatsky, Y. *et al.* Laguerre-Gaussian modes selection in diode-pumped solid-state lasers. *Opt. Rev.* **19**, 201-221 (2012).

28  Forbes, A. Controlling light's helicity at the source: orbital angular momentum states from lasers. *Phil. Trans. R. Soc. A* **375**, 1-14 (2017).

29  Naidoo, D., Ait-Ameur, K., Brunel, M. & Forbes, A. Intra-cavity generation of superpositions of Laguerre-Gaussian beams. *Appl. Phys. B* **106**, 683-690 (2012).

30  Ishaaya, A. A., Davidson, N. & Friesem, A. A. Very high-order pure Laguerre-Gaussian mode selection in a passive Q-switched Nd : YAG laser. *Opt. Express* **13**, 4952-4962 (2005).

31  Litvin, I. A., Ngcobo, S., Naidoo, D., Ait-Ameur, K. & Forbes, A. Doughnut laser beam as an incoherent superposition of two petal beams. *Opt. Lett.* **39**, 704-707 (2014).

32  Lin, D., Daniel, J. M. & Clarkson, W. A. Controlling the handedness of directly excited Laguerre-Gaussian modes in a solid-state laser. *Opt. Lett.* **39**, 3903-3906 (2014).

33  Kim, D. J. & Kim, J. W. Direct generation of an optical vortex beam in a single-frequency Nd:YVO4 laser. *Opt. Lett.* **40**, 399-402 (2015).

34  Naidoo, D. *et al.* Controlled generation of higher-order Poincaré sphere beams from a laser. *Nature Photon.* **10**, 327-332 (2016).

35  Beth, R. A. Mechanical Detection and Measurement of the Angular Momentum of Light. *Phys. Rev.* **50**, 115-125 (1936).

36  Bliokh, K. Y., Rodríguez-Fortuño, F. J., Nori, F. & Zayats, A. V. Spin–orbit interactions of light. *Nature Photonics* **9**, 796-808 (2015).

37  Marrucci, L., Manzo, C. & Paparo, D. Optical Spin-to-Orbital Angular Momentum Conversion in Inhomogeneous Anisotropic Media. *Phys. Rev. Lett.* **96**, 163905 (2006).

38  Cardano, F. *et al.* Polarization pattern of vector vortex beams generated by q-plates with different topological charges. *Appl. Opt.* **51**, C1-6 (2012).

39  Magni, V. Resonators for solid-state lasers with large-volume fundamental mode and high alignment stability. *Appl. Opt.* **25**, 107-117 (1986).

40  Fox, A. G. & Li, T. Resonant Modes in a Maser Interferometer. *Bell. Syst. Tech. J.* **40**, 453-488 (1961).

41  Kaiser, T., Flamm, D., Schroter, S. & Duparre, M. Complete modal decomposition for optical fibers using CGH-based correlation filters. *Opt. Express* **17**, 9347-9356 (2009).

42  Stalder, M. & Schadt, M. Linearly polarized light with axial symmetry generated by liquid-crystal polarization converters. *Opt. Lett.* **21**, 1948-1950 (1996).

43  Zhan, Q. Cylindrical vector beams: from mathematical concepts to applications. *Adv. Opt. Photonics* **1**, 1-57 (2009).

44  Dorn, R., Quabis, S. & Leuchs, G. Sharper focus for a radially polarized light beam. *Phys. Rev. Lett.* **91**, 233901 (2003).

45  Chen, Z., Zhang, Y. & Xiao, M. Design of a superoscillatory lens for a polarized beam. *J. Opt. Soc. Am. B* **32**, 1731-1735 (2015).

46  Barreiro, J. T., Wei, T. C. & Kwiat, P. G. Remote preparation of single-photon "hybrid" entangled and vector-polarization States. *Phys. Rev. Lett.* **105**, 030407 (2010).

47  Rafayelyan, M. & Brasselet, E. Laguerre-Gaussian modal q-plates. *Opt. Lett.* **42**, 1966-1969 (2017).

48  Devlin, R. C., Ambrosio, A., Rubin, N. A., Mueller, J. P. B. & Capasso, F. Arbitrary spin-to–orbital angular momentum conversion of light. *Science* **358**, 896-901 (2017).

49  Xie, G. D. *et al.* Experimental demonstration of a 200-Gbit/s free-space optical link by multiplexing Laguerre-Gaussian beams with different radial indices. *Opt. Lett.* **41**, 3447-3450 (2016).

50  Woerdemann, M., Alpmann, C., Esseling, M. & Denz, C. Advanced optical trapping by complex beam shaping. *Laser Photonics Rev.* **7**, 839-854 (2013).

**Supplementary Information**

**Supplementary Note 1 | Working principle of the VWP.**

The VWP can be seen as a spatially variant half-wave plate, whose fast axis rotates continuously over the area of the optic around a singularity point. Its transmission efficiency is up to 96%. The orientation of its fast axis can be expressed as:

$$\theta(\varphi) = \frac{m}{2}\varphi + \varphi_0, \tag{1}$$

where $\varphi$ is the azimuthal angle, $\varphi_0$ is the orientation of the fast axis at $\varphi = 0$ and $m$ is a positive integer determined by the VWP. Its Jones Matrix can be written as[1]:

$$M(\theta) = \begin{bmatrix} \cos 2\theta & \sin 2\theta \\ \sin 2\theta & -\cos 2\theta \end{bmatrix}. \tag{2}$$

If we apply it to a horizontally-polarized light beam, the output result can be expressed as:

$$E_V = M(\theta)\begin{pmatrix} 1 \\ 0 \end{pmatrix} = \begin{pmatrix} \cos(m\varphi + 2\varphi_0) \\ \sin(m\varphi + 2\varphi_0) \end{pmatrix}. \tag{3}$$

Supplementary Equation (3) shows that each point on the transverse plane of the output beam is linearly-polarized, but the polarization direction depends on the azimuthal angle $\varphi$. Therefore, an input linearly-polarized Gaussian mode will generate a cylindrical vector beam[2]. If we apply it to a LCP or RCP OAM mode with TC of $l_0$:

$$\begin{cases} M(\theta) \times |l_0, L\rangle = |l_0 + m, R\rangle \\ M(\theta) \times |l_0, R\rangle = |l_0 - m, L\rangle \end{cases}, \tag{4}$$

where $L$ and $R$ refer to the LCP and RCP states, both the SAM and OAM change during the process of the spin-orbital coupling. A LCP (or RCP) OAM mode with TC of $l_0$ passing through the VWP will become a RCP (or LCP) OAM mode with TC of $l_0+m$ (or $l_0-m$)[1]. When $l_0$ equals to zero (e.g., the Gaussian mode), the TC will be controlled only by the $m$ value of the VWP and the handedness of the incident circularly-polarized beam.

However, the VWP loads TC to a Gaussian mode by adding a spiral phase, which will generate a HyGG mode instead of a pure LG mode[3,4]:

$$u_{HyGG}^l = \exp\left(-\frac{r^2}{w_0^2}\right)\exp(il\varphi), \tag{5}$$

where $r$ is the radial coordinate and $w_0$ is the waist radius of the incident Gaussian mode. Expanding the HyGG mode in the LG mode representation, we rewrite Supplementary Equation (5) as:

$$u_{HyGG}^l = \sum_{p=0}^{\infty} c_l^p LG_l^p. \tag{6}$$

Here, $|c_l^p|^2$ yields the power weighting of the corresponding $LG_l^p$ mode. Supplementary Equation (6) shows that the HyGG mode is the superposition of $LG_l^p$ modes with different $p$ indices. In our laser cavity, the undesired higher-order $LG_l^p$ modes (with $p > 0$) are suppressed by the transverse mode selection of the laser cavity.

**Supplementary Note 2 | Self-reproducing process.**
As shown in Figs. 1c and 1d, a horizontally-polarized OAM mode with a TC of $l_0$ passes through the FR with its polarization direction rotating 45° clockwise. The fast axis of the QWP is oriented horizontally, so that the linearly-polarized state turns into LCP state. After passing through the VWP, the LCP state is reversed to a RCP state and an OAM mode with TC of $l_0+m$ is generated as the output mode. Both the handedness of the circularly-polarized state and the TC reverse again after the OAM mode is bounced back by the output coupler. Travelling back through the same VWP and QWP successively, the additional TC of the OAM mode is cancelled and the polarization direction has an angle of 135° with the initial polarization direction. In this process, the LCP OAM mode turns into a linearly-polarized OAM mode with a TC of $-l_0$. Twisted by the FR again, the polarization state rotates 45° clockwise and recovers to the initial horizontal direction. The reflection of the input coupler inverses the TC to its initial value of $l_0$, while keeps horizontal polarization unchanged. Such cycle repeats in the cavity. Similarly, orienting the fast axis of the QWP along the vertical axis, a LCP OAM mode with TC of $l_0-m$ will be generated.

**Supplementary Note 3 | Highly stable intensity profile at different pump powers.** To demonstrate the excellent mode stability of the output $LG_l^0$ beams, we record the intensity patterns at a propagation distance of 50 cm at different pump powers. Supplementary Figure 1a shows the intensity patterns of the $LG_1^0$ laser mode at pump powers of 0.68 W, 1.10 W, 1.33 W, and 1.50 W from left to right, respectively. The stable intensity patterns illustrate that increasing pump power has little influence on the mode purity. The output $LG_2^0$ mode is robust as well, which can be seen from Supplementary Fig. 1b. The pump powers are 1.88 W, 2.49 W, 3.10 W, and 3.60 W from left to right, respectively.

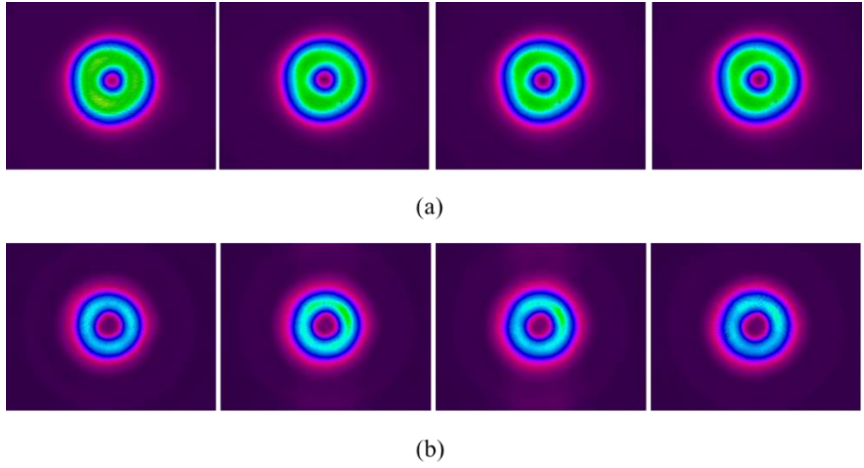

(a)

(b)

**Supplementary Figure 1. Mode stability at different pump powers.** (a) Intensity patterns of the $LG_1^0$ mode at the pump power of 0.68 W, 1.10 W, 1.33 W and 1.50 W, respectively. (b) Intensity patterns of the $LG_2^0$ mode at pump power of 1.88 W, 2.49 W, 3.10 W, 3.60 W, respectively.

**Supplementary Note 4 | Modal decomposition process.**

Modal decomposition using digital holograms is a common way to characterize laser beams[4,5]. Because all the $LG_l^p$ modes construct a complete and orthonormal basis, the output laser mode $U(x, y)$ can be expanded into the coherent superposition of the $LG_l^p$ modes,

$$U(x,y) = \sum_{l=-\infty}^{\infty} \sum_{p=0}^{\infty} c_l^p LG_l^p(x,y). \qquad (7)$$

We use the Cartesian coordinates $(x, y)$ to replace the cylindrical coordinates $(r, \theta)$ in the transverse plane normalized to propagation axis. $c_l^p$ is the modal coefficient weighting the contribution of the corresponding $LG_l^p$ mode, which can be calculated as,

$$c_l^p = \langle LG_l^p(x,y) | U(x,y) \rangle = \iint U(x,y) LG_l^p(x,y)^* dxdy. \qquad (8)$$

$LG_l^p(x, y)^*$ is the complex conjugate of $LG_l^p(x, y)$. The integration range of the inner product in Supplementary Equation (8) covers the whole transverse plane, so that Supplementary Equation (8) is impractical in reality. A Fourier transform applying on $U(x, y) \cdot LG_l^p(x, y)^*$ will simplify the calculation. We let

$$U_k(k_x, k_y) = \iint U(x,y) LG_l^p(x,y)^* \exp\left[-i(k_x x + k_y y)\right] dxdy, \qquad (9)$$

where $k_x$ and $k_y$ are wave vectors. Then the optical field on the optical axis of the Fourier plane is given by

$$U_k(0,0) = \iint U(x,y) LG_l^p(x,y)^* dxdy. \qquad (10)$$

By measuring the intensity of $U_k(0,0)$, the power weighting of the corresponding $LG_l^p$ mode can be determined to be

$$\left|c_l^p\right|^2 = \left|U_k(0,0)\right|^2. \tag{11}$$

Supplementary Equation (11) provides a practical way to determine the mode purity of the output $LG_l^p$ mode.

We use a reflective phase-only spatial light modulator (SLM, LUTO-NIR-011, HOLOEYE Corporation) to assist with the modal decomposition. The SLM can be loaded on well-designed computer-generated holograms (CGH) to generate arbitrary complex field. We use the type-3 method of complex-amplitude modulation reported by Arrizon et al. to program the phase-only CGH[6]. A high-frequency phase carrier is added into the CGH, so that the desired 1st-order diffraction beam, which recovers the conjugate of the $LG_l^p$ mode, separates from the other undesired modes in the Fourier plane[7].

Supplementary Figure 2 shows the schematic setup for the modal decomposition process according to Supplementary Equations (10) and (11). Because the SLM only works for linearly-polarized light, we turn the output circularly-polarized LG mode into a linearly-polarized one using a quarter-wave plate. It passes through a 50/50 beam splitter (BS) and is incident onto the SLM. The reflected light field carrying the information of $U(x, y) \cdot LG_l^p(x, y)^*$ propagates back to the BS. Half of this reflected light field by the BS is focused by a lens with a focal length of 75 mm to perform the Fourier transform. At the Fourier plane, the 1st-order diffraction beam recovering the field $U_k(k_x, k_y)$ is picked out by an iris and imaged on the LBP by a 50X objective. Recording the central intensities of the images, the power weightings can be determined. To measure the modal distribution of the generated $LG_1^0$ laser beam, we load a group of $LG_l^{p\,*}$ modes, $p$ varying from 0 to 5 and $l$ from 0 to 2, onto the SLM. The corresponding 1st-order diffraction patterns are recorded by the LBP (Supplementary Fig. 3). According to Supplementary Equation (11), we choose 4×4 pixels in the center of each pattern to calculate the power weighting of each $LG_l^p$ mode. The modal distribution of the $LG_2^0$ beam is measured in the similar way using pure $LG_l^p$ modes with $p = 0$ to 5 and $l = 1, 2, 3$. The results are shown in Fig. 4.

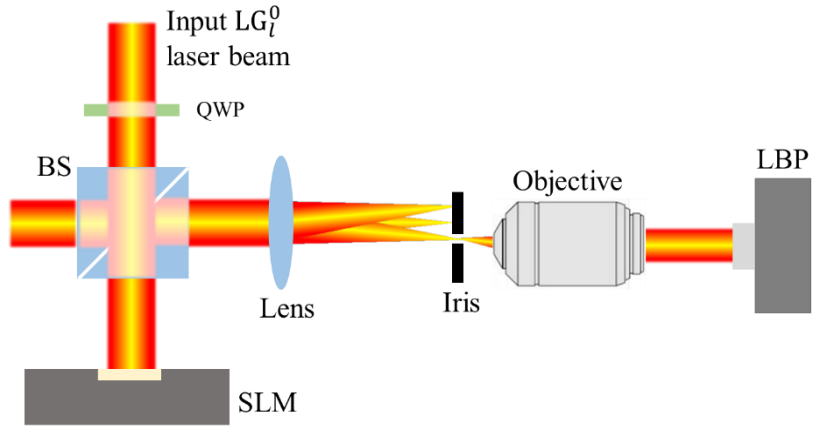

**Supplementary Figure 2. The schematic setup for modal decomposition.** The output $LG_l^0$ laser beam passes through a quarter-wave plate (QWP) and becomes horizontally polarized. Half of its light field transmits through a 50/50 beam splitter (BS) and is incident onto the spatial light modulator (SLM). After modulated by the SLM, the light field propagates backward and half of its light field reflected by the BS is focused by a lens. The 1$^{st}$-order diffraction beam is picked out by an iris and imaged by an objective onto the laser beam profile (LBP).

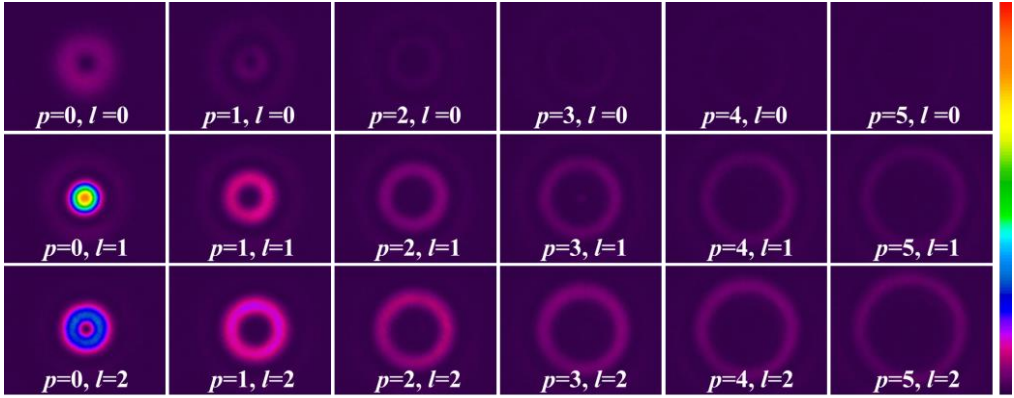

**Supplementary Figure 3. Modal decomposition patterns.** 1$^{st}$-order diffraction patterns when the input $LG_1^0$ laser beam is modulated by the SLM carrying different pure $LG_l^{p\,*}$ modes with $p = 0$ to 5 and $l = 0, 1, 2$, respectively.

**Supplementary Note 5 | Experimental setup for laser output of cylindrical vector beams.**

The experimental setup is shown in Supplementary Fig. 4. An a-cut 0.5 at% Nd-doped YVO$_4$ crystal with its dimensions of 3 mm × 3 mm × 8 mm serves as the gain medium. One of its ends, as the input coupler, is coated with reflectivity of >99.9% at 1064 nm and transmittance of >97.7% at 808 nm; the other end has a high transmittance of >99.8% at 1064 nm. The crystal is mounted on a water-cooled copper holder maintained at the room temperature and oriented to generate horizontally-polarized light. The output coupler is a plane mirror with a transmittance of 10% at 1064 nm. The lens with focal length $f$ is inserted in the cavity to keep the cavity satisfying the stability condition. The pump source is a fiber-coupled laser diode outputting a continuous-wave at 808 nm wavelength, which is coupled out by a telescope and forms a 200 μm-in-diameter focusing spot in the crystal. A vortex half-wave plate (VWP) fixed on a rotator is inserted into the laser cavity for producing cylindrical vector beams. The PBS following the crystal sets the incident light into horizontal polarization and eliminates the unwanted vertically-polarized light caused by imperfect polarization modulation of the VWP. The polarization distribution of the generated cylindrical vector beam can be controlled by replacing VWP with different $m$ and/or rotating the VWP. The generated cylindrical vector beams using the VWP of $m = 1$ ($m = 2$) are under the cavity parameters of $f = 200$ mm, $L_A = 205$ mm and $L_B = 45$ mm ($f = 400$ mm, $L_A = 380$ mm and $L_B = 45$ mm).

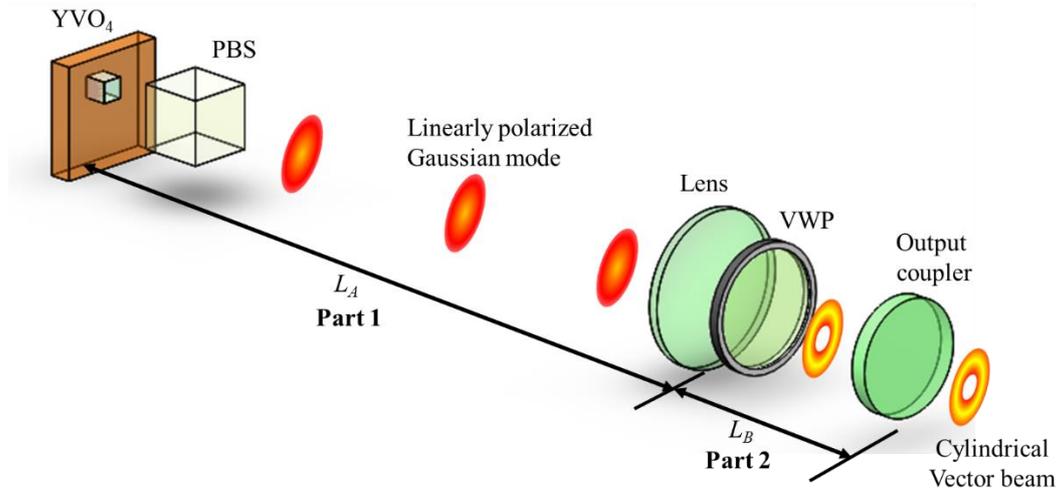

**Supplementary Figure 4. Experimental setup for laser output of the cylindrical vector beams.** An end-pumped solid-state laser comprises of a Nd:YVO$_4$ crystal, a lens and an output coupler working at the wavelength of 1064 nm. The Nd:YVO$_4$ crystal is pumped by an 808 nm-wavelength fiber-coupled diode laser. Following the Nd: YVO$_4$ crystal, a polarization beam splitter (PBS) and a vortex half-wave plate (VWP) are placed in the cavity successively. All of them are antireflection coated and oriented correctly, so that the cavity satisfies light reversibility condition.

**Supplementary Note 6 | Numerical simulations.**
Based on the Fox-Li method, an optical cavity can be expanded into a transmission line by replacing cavity mirrors with equivalent lenses[8]. For simplifying the calculation process, we only calculate the scalar cavity mode in a passive optical cavity, in which the polarization transformation, gain and loss are ignored. A pinhole is needed to filter out undesired higher-order radial modes. We set it on the output coupler because the distance between them is very small in the experimental setup. The distance between the lens and the VWP is neglected as well. Besides, the simplified model of scalar field allows us to replace the VWP with two spiral phase plates (SPPs) of opposite TCs. Therefore, one round-trip path of our laser cavity is equivalent to a complete cycle shown in Supplementary Fig. 5. A scalar light field starting from the front mirror travels a distance of $L_A$. It passes through the lens and the SPP with TC of $l$, and then reaches the pinhole after propagating a distance of $L_B$. A fraction of light is blocked by the pinhole while the majority propagates a distance of $L_B$. After passing through the other SPP with a TC of $-l$ and another lens, the spiral phase front is cancelled out. The light beam travels a distance $L_A$ to the front mirror and finishes one round trip. The light beam repeats the cycle until a stable cavity mode is formed. The iterative procedure described above is calculated step by step by Matlab programming. Supplementary Figure 6a shows the simulated intensity pattern of the output $LG_1^0$ beam propagating a distance of 30 cm, which is nearly the same as the experimental result shown in Supplementary Fig. 6b. Furthermore, the intensity profiles of the stable intracavity mode at different locations have been also calculated. Supplementary Figure 7 presents intensity profiles at the distances of 200 mm, 150 mm, 100 mm, 50 mm, and 0 mm away from the front mirror, respectively, which show that the mode in $L_A$ evolves from a donut-intensity pattern on the lens into a Gaussian pattern on the front mirror.

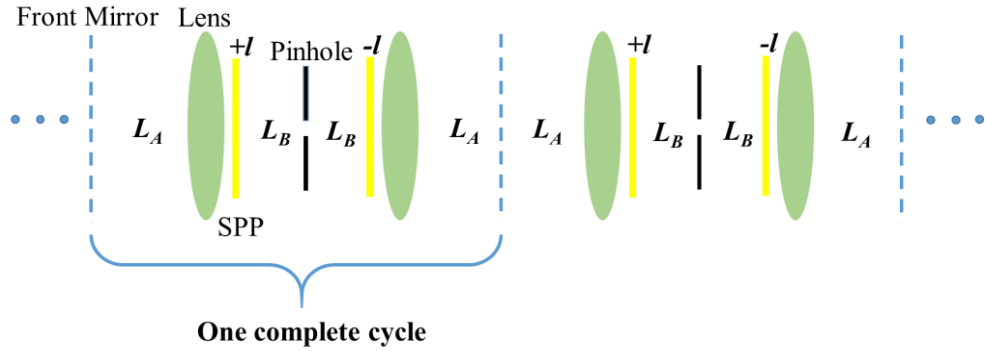

**Supplementary Figure 5. Numerical simulation process.** A light beam starting from the front mirror propagates through the lens, SPP with TC of $+l$, pinhole, SPP with TC of $-l$, and lens successively to finish one complete cycle, and repeats the cycle until the stable cavity mode is formed.

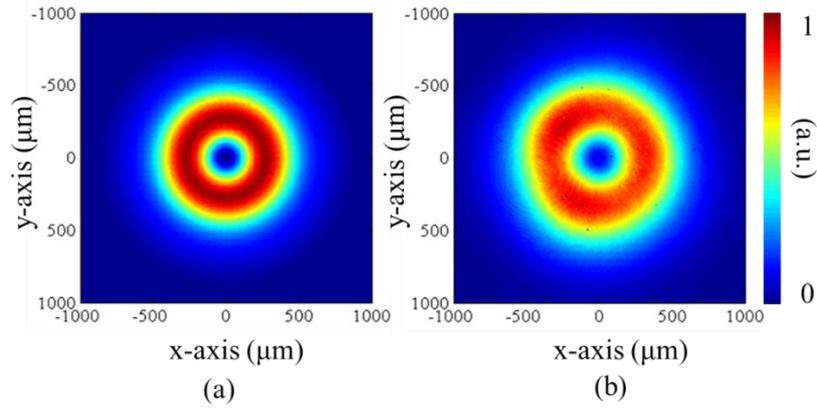

**Supplementary Figure 6. The comparison between simulated and experimental results.** The simulated (a) and experimental (b) intensity patterns of the output $LG_1^0$ laser mode at the distance of 30 cm away from the output coupler.

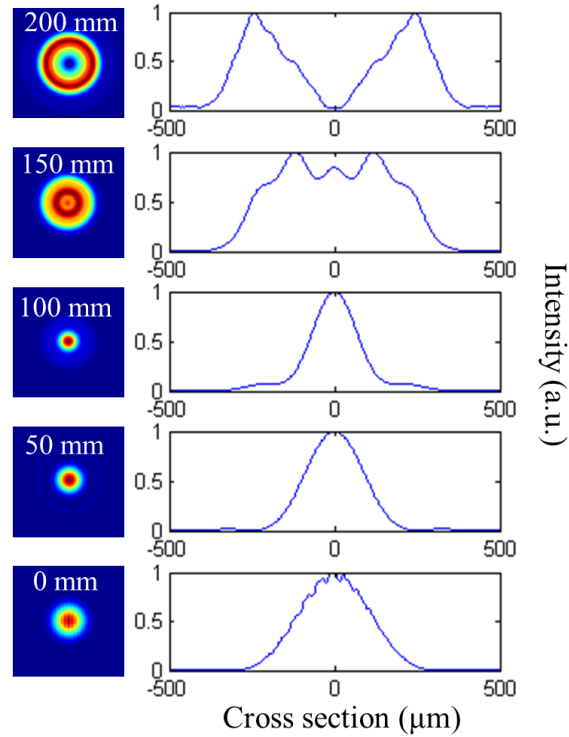

**Supplementary Figure 7. The simulated intensity profiles inside the cavity.** From top to bottom are intensity profiles at the distances of 200 mm, 150 mm, 100 mm, 50 mm, 0 mm away from the front mirror, respectively.

**Supplementary References**

1      Marrucci, L., Manzo, C. & Paparo, D. Optical Spin-to-Orbital Angular Momentum Conversion in Inhomogeneous Anisotropic Media. *Phys. Rev. Lett.* **96**, 163905 (2006).

2      Cardano, F. *et al.* Polarization pattern of vector vortex beams generated by q-plates with different topological charges. *Appl. Opt.* **51**, C1-6 (2012).

3      Karimi, E., Zito, G., Piccirillo, B., Marrucci, L. & Santamato, E. Hypergeometric-gaussian modes. *Opt. Lett.* **32**, 3053-3055 (2007).

4      Sephton, B., Dudley, A. & Forbes, A. Revealing the radial modes in vortex beams. *Appl. Opt.* **55**, 7830-7835 (2016).

5      Kaiser, T., Flamm, D., Schroter, S. & Duparre, M. Complete modal decomposition for optical fibers using CGH-based correlation filters. *Opt. Express* **17**, 9347-9356 (2009).

6      Ando, T., Ohtake, Y., Matsumoto, N., Inoue, T. & Fukuchi, N. Mode purities of Laguerre-Gaussian beams generated via complex-amplitude modulation using phase-only spatial light modulators. *Opt. Lett.* **34**, 34-36 (2009).

7      Arrizon, V., Ruiz, U., Carrada, R. & Gonzalez, L. A. Pixelated phase computer holograms for the accurate encoding of scalar complex fields. *J. Opt. Soc. Am. A* **24**, 3500-3507 (2007).

8      Fox, A. G. & Li, T. Resonant Modes in a Maser Interferometer. *Bell. Syst. Tech. J.* **40**, 453-488 (1961).